\documentclass[prb,twocolumn,showpacs,showkeys,superscriptaddress]{revtex4-1}
\usepackage{graphicx}
\usepackage{times}                 
\usepackage{amsmath}
\usepackage{amsfonts}
\usepackage{amssymb}
\usepackage{times}
\usepackage{xcolor}
\date{\today}

\begin{document}
	\title[Magnetic skyrmions in cylindrical nanostructures]
	{Magnetic skyrmions in cylindrical ferromagnetic nanostructures with chiral interactions}
	\author{D.~Kechrakos}
	\email{dkehrakos@aspete.gr}
	\affiliation{Department of Education, School of Pedagogical and Technological Education, Athens, GR-15122 }
	\author{L.~Tzannetou}
	\affiliation{Core Department, National and Kapodistrian University of Athens, Evripos Campus, Evia, GR-34400}
	\author{A.~Patsopoulos}
	\affiliation{Department of Physics, National and Kapodistrian University of Athens, Athens, GR-15784}
	\keywords{magnetic skyrmions; Dzyaloshinskii-Moriya interactions; Monte Carlo; nanotubes\\
	DOI: 10.1103/PhysRevB.102.054439}
		
	\begin{abstract}
	We study the geometrical conditions for stabilizing magnetic skyrmions in cylindrical nanostrips and nanotubes of ferromagnetic materials with chiral interactions.
	We obtain the low-temperature equilibrium state of the system implementing a simulation annealing technique for a classical spin Hamiltonian with competing isotropic exchange and chiral interactions, radial anisotropy and an external field.
	We address the impact of surface curvature on the formation, the shape and the size of magnetic skyrmions.
	We demonstrate that the evolution of the skyrmion phase with the curvature	is controlled by the competition between two characteristic lengths, namely the curvature radius, $R$ (geometrical length) and the skyrmion radius, $R_{sk}$ (physical length). 
	In narrow nanotubes ($R<R_{sk}$) the skyrmion phase evolves to a stripe phase, while in wide nanotubes ($R>R_{sk}$) a mixed skyrmion-stripe phase emerges.
	Most interestingly, the mixed phase is characterized by spatially separated  skyrmions from stripes owing to the direction of the applied field relative to the surface normal.
	In the stability regime ($R \gtrsim R_{sk}$) skyrmions remain circular and preserve their size as a consequence of their topological protection. 
	Zero-field skyrmions are shown to be stable on curved nanoelements with free boundaries within the same stability region ($R\gtrsim R_{sk}$).
	The experimental and technological perspectives from the stability of skyrmions on cylindrical surfaces are discussed. 
	\end{abstract}
	\maketitle
\section{Introduction}
\label{sec:intro}
Magnetic skyrmions are self-localized vortex-like spin structures with axial symmetry. \cite{bog94a}
They have been mainly studied in non-centrosymmetric bulk crystals and their thin films,\cite{muhl09,pap09,yux10} as well as, in ultrathin ferromagnetic (FM) films on heavy metal (HM) substrates,\cite{hein11,rom13} in which a sizable Dzyaloshinskii-Moriya interaction (DMI) \cite{dzi58,mor60}  induces their stability.
From the point of view of technological applications, two-dimensional magnetic skyrmions formed in FM-HM interfaces have potentials for a variety of innovative robust and high-density magnetic storage technologies due to their protected topology and nanoscale size.\cite{fer13} 
Magnetic skyrmions on FM-HM nanostrips can be driven by a transverse spin current,\cite{fer13,samp13,nag13} that is generated by a longitudinal electrical current with five to six orders of magnitude smaller density than that needed to electrically drive a typical domain wall,\cite{rom13} thus pointing to energy efficient skyrmion-based racetrack-type memory devices.\cite{fer13, par08} 
However, current-driven skyrmions will drift towards the racetrack side edges due to the action of a magnetic Magnus force stemming from the chirality of their spin structure.\cite{iwa13,yux12}
This phenomenon known as the Skyrmion Hall effect (SkHE) leads to their annihilation at the racetrack edge and the loss of stored information. 
Various proposals\cite{zha16,bar16,pur16,lai17,foo15} for creating a potential barrier to the sideways drift of skyrmions have been presented, which aim to the confinement of skyrmions in the central region of the racetrack.
Tuning of the perpendicular \cite{foo15} or the crystalline \cite{lai17} magnetic anisotropy, transverse modulation of the ferromagnetic layer thickness \cite{pur16} and transverse modulation of the ferromagnetic damping constant\cite{liu16} have been proposed as methods to create a low resistance path for skyrmions in the middle part of a nanostrip and suppression of the SkHE.
Along the same spirit, an exchange coupled pair of skyrmions hosted in an antiferromagnetically coupled pair of nanostrips were shown to exhibit null SkHE,\cite{zha16b} due to their opposite chiralities.
Synthetic antiferromagnets are promising candidate systems for realization of null SkHE, however their requirement for double amount of material raises a practical issue in device design.
A final aspect hampering the use of magnetic skyrmions in racetrack memory applications, is their uncontrollable excitation realized at the free side edges of nanostrips and thin films\cite{ran17} leading to error reading-writing events.

From the above, it appears that the possibility of magnetic skyrmions generation and manipulation on boundary-free samples would be a desirable direction of research and curved nanostructures, as for example,  magnetic nanowires and nanotubes, constitute a promising option.
The magnetic structure and soliton-type excitations on curvilinear nanostructures have attracted intensive research effort in recent years, motivated by the fact that the curvilinear geometry and topology of a nanostructure offer a tool for tailoring the magnetic state of a FM sample.\cite{streu16} 
The appearance of curvature-induced DMI\cite{pyl15} in curved ferromagnetic thin films offers the possibility to form small-sized skyrmions in the region of maximal curvature.
This general result was thoroughly studied for spherical ferromagnetic shells with and without intrinsic DMI.\cite{kra16}. 
Also, the skyrmion radius is controlled by the curvature gradient, which results in tunable-size skyrmions.\cite{pyl15}  
Besides, multiplet of skyrmion states with possible switching
between them and reconfigurable skyrmion lattices have been realized in the region of a curvilinear defect on an otherwise planar surface.\cite{kra18}
For the particular case of cylindrical nanotubes, recent numerical works\cite{wan19,huo19} study  the statics and current-driven dynamics of Bloch skyrmions in the absence\cite{wan19} or presence\cite{huo19} of an applied magnetic field.
These works report a weakly elongated shape of skyrmions hosted on nanotubes.\cite{huo19,wan19} 
Also, the skyrmion size increases weakly with nanotube radius\cite{huo19}.
When a uniform field is applied normal to the nanotube axis there exists a critical polar angle beyond which the current-driven skyrmion deforms and annihilates. The critical angle is independent on nanotube radius and decreases with increasing field strength.\cite{huo19}
In previous numerical studies\cite{huo19} the effect of film curvature on the characteristics of skyrmions (shape, size) has been considered under dynamic conditions, namely during the electric current-driven motion, which could in principle, interfere with the skyrmion structure.

In the present work, we focus on the low-temperature equilibrium properties of the skyrmion phase in cylindrical ferromagnetic nanostructures with chiral  interactions (DMI) and examine the conditions under which curvature-driven skyrmion instability occurs. 
We study N\'{e}el Skyrmions, as those formed on a thin ferromagnetic film on a heavy metal substrate, because the FM/HM interfaces have been  so far most promising from the point of view of technological exploitation in skyrmion-based devices\cite{fer13}.
Our structural model accounts for adaption of the DMI vector to the curvature of the nanostructure, thus providing a more realistic description of the interplay between isotropic exchange (Heisenberg) and chiral interactions on curved surfaces.
We focus on the interplay between intrinsic interactions (exchange, DMI, anisotropy) under increasing curvature using a lattice spin model.
Thus, we do not include in the total energy the curvature-induced DMI and curvature-induced anisotropy terms, as introduced for a general curved surface in Ref.\cite{kra16}.  
More specifically, for the cylindrical geometry considered here, the curvature-induced terms have interaction strengths expressed as\cite{kra16,wan19} 
$D^{curv}=2A/R$ and $K^{curv}=A/R^2$, 
with $A$ the exchange stiffness and $R$ the cylinder radius. 
For parameters corresponding to a typical FM/HM interface \cite{hag15,lel19} and large enough cylinder radius that concern us here (see Section~\ref{sec:results}), the curvature-induced parameters become $5$ to $10$ times smaller than the corresponding intrinsic parameters and have thus been ignored.   

Our results demonstrate the feasibility of skyrmion formation along the ridge of a cylindrical nanotube, where the external field remains almost normal to the surface, provided that the radius of the nanotube remains at least comparable to the skyrmion radius $(R \gtrsim R_{sk} )$. 
Skyrmion instability is associated with a decrease of the radial component of the applied field below a critical value or equivalently with a critical curvature angle ($\phi_0$) for a given applied field. 
This effect leads to shrinkage of the skymion-phase pocket in the anisotropy-field phase diagram.
Shape analysis of the hosted skyrmions shows that the circular shape is approximately preserved up to the instability point.
The same geometrical criteria ($\phi_0 \ge \phi_0^c, R \gtrsim R_{sk} $) define the stability regime of zero-field skyrmions on curved nanoelements.
\section{Lattice Model and Simulation Method}
\label{sec:model}
\begin{figure}[htb!]
	\centering
	\includegraphics[width=0.95\linewidth]{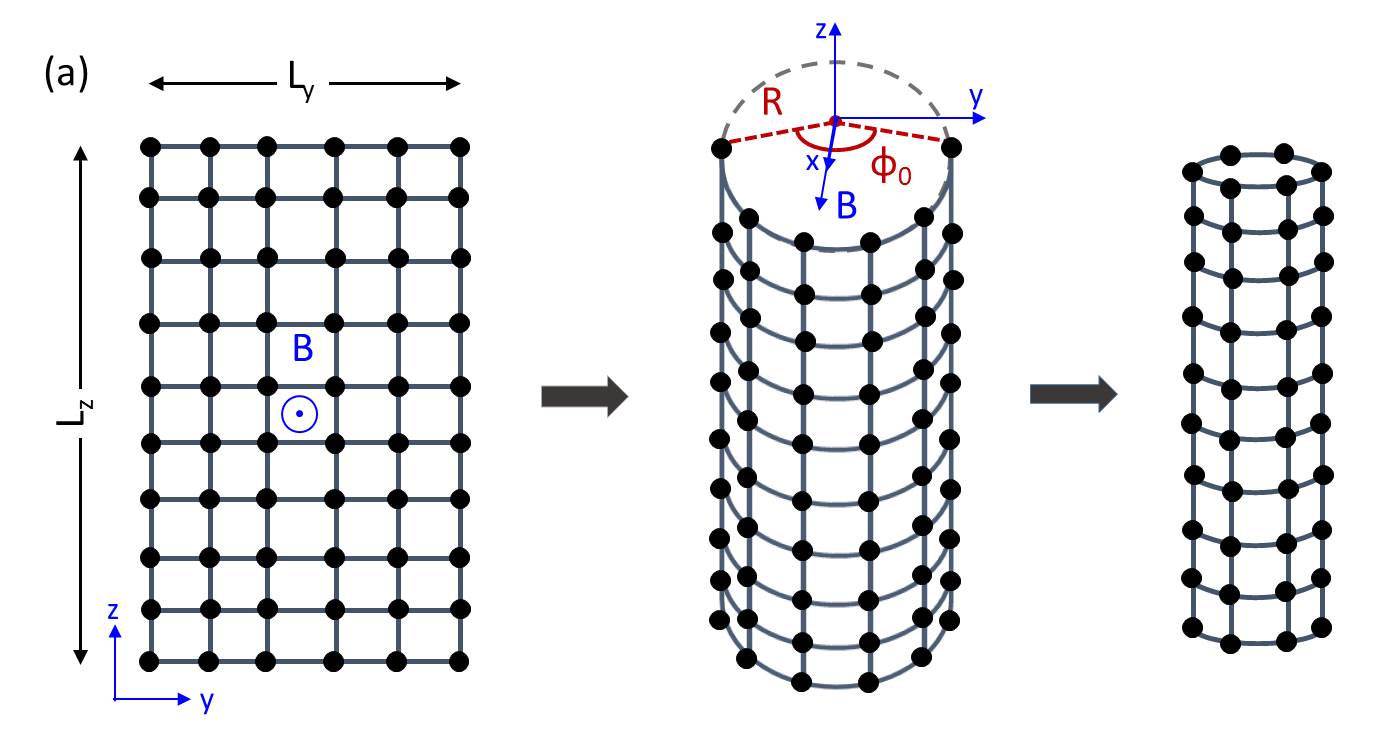}
	\includegraphics[width=0.95\linewidth]{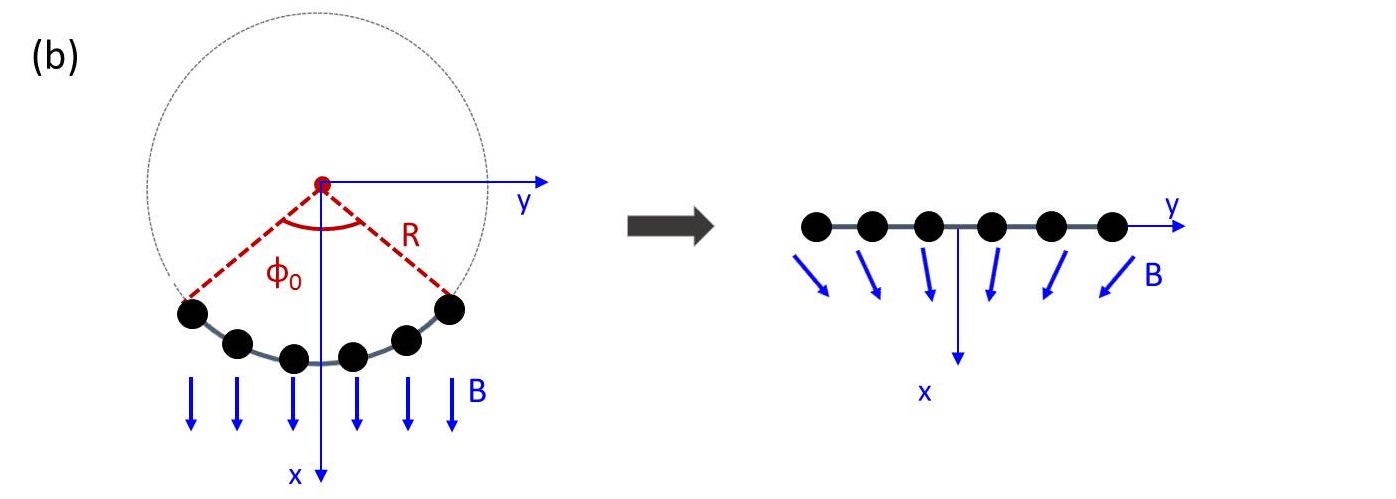}
    \caption{
		 (a) A planar nanostrip, modeled by a square lattice with length $L_z=10a$ and width $L_y=6a$,  is wrapped around cylinders with successively smaller radius and eventually forms a cylindrical nanotube. Every cylinder has the main axis along the Cartesian $z$-axis and is characterized by the curvature radius $R$ and the central angle $\phi_0$. The $x$-axis is normal to the cylindrical surface at the midpoint of the nanostrip.
		 (b) Top view of a curved nanostrip with $L_y=6a$ under a uniform magnetic field along the $x$-axis (left) as seen in unfolded view (right).
	}
	\label{fig:sketch}
\end{figure}

We consider a planar nanostrip in the $yz$-plane, cut from a two-dimensional square lattice with lattice constant $a$ and Castesian coordinates  $-L_y/2 \le y \le L_y/2$ and $-L_z/2 \le z \le L_z/2$. 
A curved nanostrip is formed by wrapping the initial nanostrip around a cylinder of radius $R$ along the $z$-axis.
The principal direction [01] of the square lattice is always parallel to the $z$-axis. 
The $x$-axis is always normal to the cylindrical surface at the midpoint of the nanostrip (Fig.\ref{fig:sketch}).
The width of the nanostrip defines the central angle of the curved nanostrip through $\phi_0=L_y/R$. 
A planar nanostrip ($R\rightarrow\infty,\phi_0=0 $) and a cylindrical nanotube ($R\ne0,\phi_0=360^0$) are then considered as limiting cases of the curved nanostrip.
The geometry of our two-dimensional lattice model approximately describes a continuous cylindrical nanostrip with infinitely small thickness ($t\ll R$).
We are interested in the magnetic behavior of a curved interface between a thin ferromagnetic film and a heavy metal substrate.
For this purpose we use the following expression for the total energy 
\begin{eqnarray} 
E = 
-\frac{1}{2} J \sum_{<ij>} \textbf{m}_i \cdot \textbf{m}_j 
\nonumber \\
-\frac{1}{2} d \sum_{<ij>} \textbf{D}_{ij} \cdot 
(\textbf{m}_i \times \textbf{m}_{j})
\nonumber \\
-k \sum_i (\textbf{m}_i \cdot \textbf{n}_i )^2
-h \sum_i \textbf{m}_i \cdot \textbf{h}_i,
\label{eq:energy}
\end{eqnarray}
where            
$\textbf{m}_i$ is the magnetic moment unit vector (spin) of $i$-th site.
The first term in Eq.(\ref{eq:energy}) is the symmetric exchange energy contribution and is restricted to first nearest neighbor sites denoted as $<ij>$. 
The second term is the antisymmetric exchange (DMI) energy.  
The DMI vector $\textbf{D}_{ij}$  lies on the surface of the nanostrip with direction normal to the first nearest neighbor bond vector $\textbf{r}_{ij}$ and has the form   
$\textbf{D}_{ij}=\textbf{n}_i \times \textbf{r}_{ij}$,
with $\textbf{n}_i$ the unit vector in the radial direction on site $i$.
This expression is analogous to 
$\textbf{D}_{ij}=\textbf{x} \times \textbf{r}_{ij}$
that describes the DM coupling at planar interfaces along the $yz$-plane.\cite{hag15,yin16}
The main difference with the planar case is that for curved nanostrips the direction of the vector $\textbf{D}_{ij}$ becomes site-dependent, owing to the variation of the radial direction across the surface. 
The consequences of this geometrical condition are discussed in the next section.
The $1/2$ prefactor of the first and second terms accounts for the double-counting of energy contribution from pairs of nearest neighboring sites.
The third term is the uniaxial anisotropy energy with easy axis along the local radial direction.
We assume here a generalization of the perpendicular anisotropy observed in thin ferromagnetic films on a heavy metal substrate\cite{fer13,hag15}.
The final term is the Zeeman energy due to an applied field, which is assumed either homogeneous along the $x$-axis ($\textbf{h}_i=\textbf{x}$) or radial ($\textbf{h}_i=\textbf{n}_i$), as explicitly mentioned in the numerical results. 
Magnetostatic energy terms are neglected in Eq.(\ref{eq:energy}), because for infinitely thin shells ($t \ll R$) they can be reduced to a correction to the local anisotropy. \cite{roh13,sla05}  
The energy parameters $J,d,k,h$ entering Eq.(\ref{eq:energy}) are related to the corresponding micromagnetic parameters through the relations
$J\approx 2Aa, 
d \approx Da^2$, and 
$k \approx K_ua^3$,  
$h\approx M_sBa^3$,
where $A$ is the exchange stiffness, $D$ the DMI energy density, $K_u$ the anisotropy energy density, $M_s$ the saturation magnetization and $B$ the applied field. 
The above relations are exact in the case of cubic discretization of a planar nanostrip.
However, we use them also in the case of a cylindrical nanostrip assuming that the discretization cell is almost cubic, which is a reasonable approximation when $R \gg a$. 
We use material parameters typical of a transition metal thin film on a heavy metal substrate,\cite{hag15,lel19} namely
$M_s=580kA/m$, 
~$A=10pJ/m$, 
~$D=5mJ/m^2$ and 
~$K_u=500kJ/m^3$.
The applied field is $B=0.9~T$ and the lattice constant $a=2nm$, which is below the magnetic length $l_m=\sqrt{A/K_u}\approx 4.5nm$.
Then the rationalized (dimensionless) parameters read
$
d/J=0.5, 
k/J=0.1, 
h/J=0.1
$
and they consist a complete set of parameters that determines the equilibrium state configuration of the spin system.

In the absence of anisotropy ($k=0$) the pitch length of the helical phase\cite{kee15,sek16}  
$p=2\pi a / tan^{-1}(d/J)$ 
serves as a rough estimate of the 2D skyrmion diameter.\cite{moc12} 
For the material parameters mentioned above we obtain $p\approx 13.6a$. 
The anisotropy ($k\ne 0$) introduced in our model  is expected to weakly reduce the skyrmion radius.
We use the value of the pitch length as a rough estimate of the skyrmion size in anisotropic samples and the ratio $p/L$ as an estimate of the role of finite size effects, that is useful when we change the discretization level (see Section~\ref{sec:Sk_PD}).\cite{kee15} 

To obtain the low temperature equilibrium state we follow the simulated annealing method\cite{kir83} using the Metropolis Monte Carlo algorithm with single spin updates and temperature-dependent spin aperture that accelerates the approach to equilibrium. 
In particular, we follow a field-cooling (FC) protocol  under an applied field $h/J=0.1$ bringing the system from the high- temperature ($k_BT/J=20$) demagnetized state to the low-temperature ($k_BT/J=10^{-3}$) state, with a variable temperature step $dT/T=5\%$, which allows for longer relaxation periods and as the temperature drops.
At each temperature step we perform $5000$ Monte Carlo steps per spin (MCSS) for thermalization followed by $5000$ MCSS for calculations of thermodynamic quantities.  
The thermal averages of macroscopic quantities  are obtained from sampling every $\tau=10$ MCSS, in order to minimize statistical correlations between sampling points. 
The thermodynamic quantities at each temperature are averaged over $N_{seq}=50$ independent relaxation sequences to obtain the statistical errors.
\section{Results and Discussion}
\label{sec:results}
\subsection{Skyrmion phase}
\label{sec:Sk_multi}
\begin{figure} [htb!]
	\centering
	\includegraphics[width=0.95\linewidth]{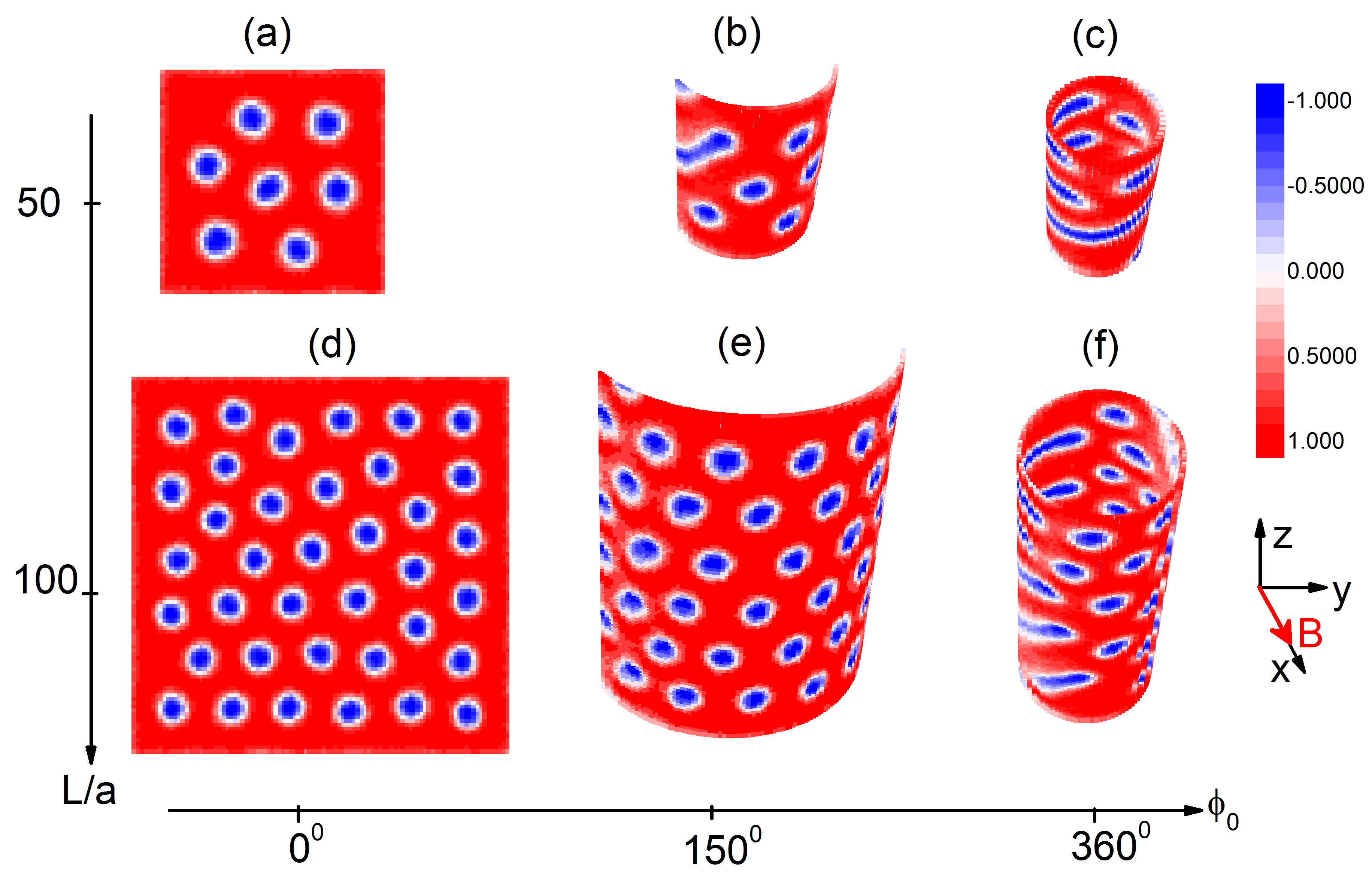}
	\caption{
		(Color online) Ground state configuration showing skyrmion formation in cylindrical nanostructures along the $z$-axis under application of a uniform magnetic field along the $x$-axis. 
		Cylindrical surfaces are constructed by gradually wrapping an initial square sample $L_y=L_z\equiv L$. 
		Spin configurations are color coded according to the value of magnetization along the applied field direction (red=+1, blue=-1).
		(a), (d) planar surfaces,
	    (b) $L=50a,  \phi_0=150^0, R=19.1a$,
	    (c) $L=50a,  \phi_0=360^0, R=8.0a$,
	    (e) $L=100a, \phi_0=150^0, R=38.2a$, and
	    (f) $L=100a, \phi_0=360^0, R=15.9a$,
	    with $a=2nm$.
	    With increasing angle of curvature ($\phi_0$) the skyrmion phase transforms to either a spiral phase, as in (c), or to a mixed skyrmion-spiral phase, as in (f), depending on the value of the curvature radius ($R$). 
	    Parameters: $d/J=0.5, k/J=0.1, h/J=0.1$ and $k_BT/J=10^{-3}.$
    }
	\label{fig:Skphase}
\end{figure}
We consider first the evolution of the skyrmion phase as the curvature of the nanostructure increases. 
We start from a planar surface in the yz-plane and fold it gradually to form an open cylindrical surface with axis along the Cartesian $z$-axis and eventually, a closed cylindrical surface corresponding to a nanotube (Fig.\ref{fig:sketch}).
When we curve the 2D sample, we preserve the dimensions $(L_y,L_z)$ of the initial planar system in order to emphasize the role of curvature and exclude finite size effects.
Periodic boundary conditions are used solely along the z-axis of our curved samples, except for nanotubes, when the lateral free boundaries couple among themselves, naturally.

As the curvature of the nanostrip increases the low temperature magnetic state is modified. In Fig.\ref{fig:Skphase} we show that for a planar systems the well-known skyrmion lattice\cite{yis09} occurs, which consists of a hexagonal arrangement of skyrmions.
Obviously, the number of skyrmions increases with the area of the planar sample, however their spatial density remains almost unchanged.

As the angle of curvature  increases, skyrmions close to the free edges of the curved surface become elongated and finally transform into spirals.
This effect becomes more evident in smaller samples, which are characterized by smaller values of the curvature radius, as in Fig.\ref{fig:Skphase}(b),(c).
In a small nanotube with radius $R=8a$ (Fig.\ref{fig:Skphase}(c)) stripes form almost all around the surface.
On the contrary, in a larger nanotube with radius $R=15.9a$ isolated skyrmions are observed along the front and the back ridge of the cylinder, where the external field is almost normal to the surface, but spiral structures form along the left and right sides of the large tube (Fig.\ref{fig:Skphase}(f)) where the applied field is almost tangential to the surface.

Thus, skyrmion formation on nanotubes is strongly dependent on the nanotube radius, with large radius nanotubes supporting the coexistence of both skyrmion and stripe phases.
We underline the fact that the two phases are spatially separated with skyrmions forming along the ridge and stripes forming on the sides of the nanotube. 
The width of the region supporting skyrmions
is determined by the size of the skyrmion radius ($R_{sk}$) relative to the curvature radius ($R$). 
This point is discussed further below.

To quantify the evolution of the skyrmion phase with sample curvature, as depicted in Fig.\ref{fig:Skphase}, we calculate the topological charge ($Q$).
For a three component spin field $\textbf{m}(\phi,z)$ on a cylindrical surface described by the coordinates ($\phi,z$), the topological charge is given as \cite{kra16}
\begin{eqnarray}
Q=\frac{1}{4\pi} \iint d\phi~dz~
\textbf{m}\cdot
(\frac{\partial\textbf{m}}{\partial\phi}
\times
\frac{\partial\textbf{m}}{\partial z}).
\label{eq:topol_charge}
\end{eqnarray}
For the numerical computation we implement an discrete form of the topological charge\cite{ber81} appropriate to a square lattice wrapped around a cylindrical surface. 
Skyrmions have a topological charge $Q=\pm1$, depending on the direction of the applied field relative to the surface normal. 
Thus the absolute value of $Q$ for a nanostrip in the skyrmion phase is equal to the number of skyrmions supported.

For planar nanostrips shown in Fig.\ref{fig:Skphase}a ($Q=7.2$) and Fig.\ref{fig:Skphase}d ($Q=37.8$), the values of $Q$ deviate weakly from integer values due to the misalignment of the moments located on the free boundaries of the sample\cite{roh13} and the thermal fluctuations inherent to the Monte Carlo method.
For curved surfaces, however, the shape distortion of the skyrmions and their  evolution to stripe-like structures is not characterized by integer values of $Q$, thus the calculation of $Q$ based on Eq.(\ref{eq:topol_charge}) assumes non-integer values and is only indicative of the number of skyrmions observed in the mixed phase.

\begin{figure} [htb!]
	\centering
	\includegraphics[width=0.95\linewidth]{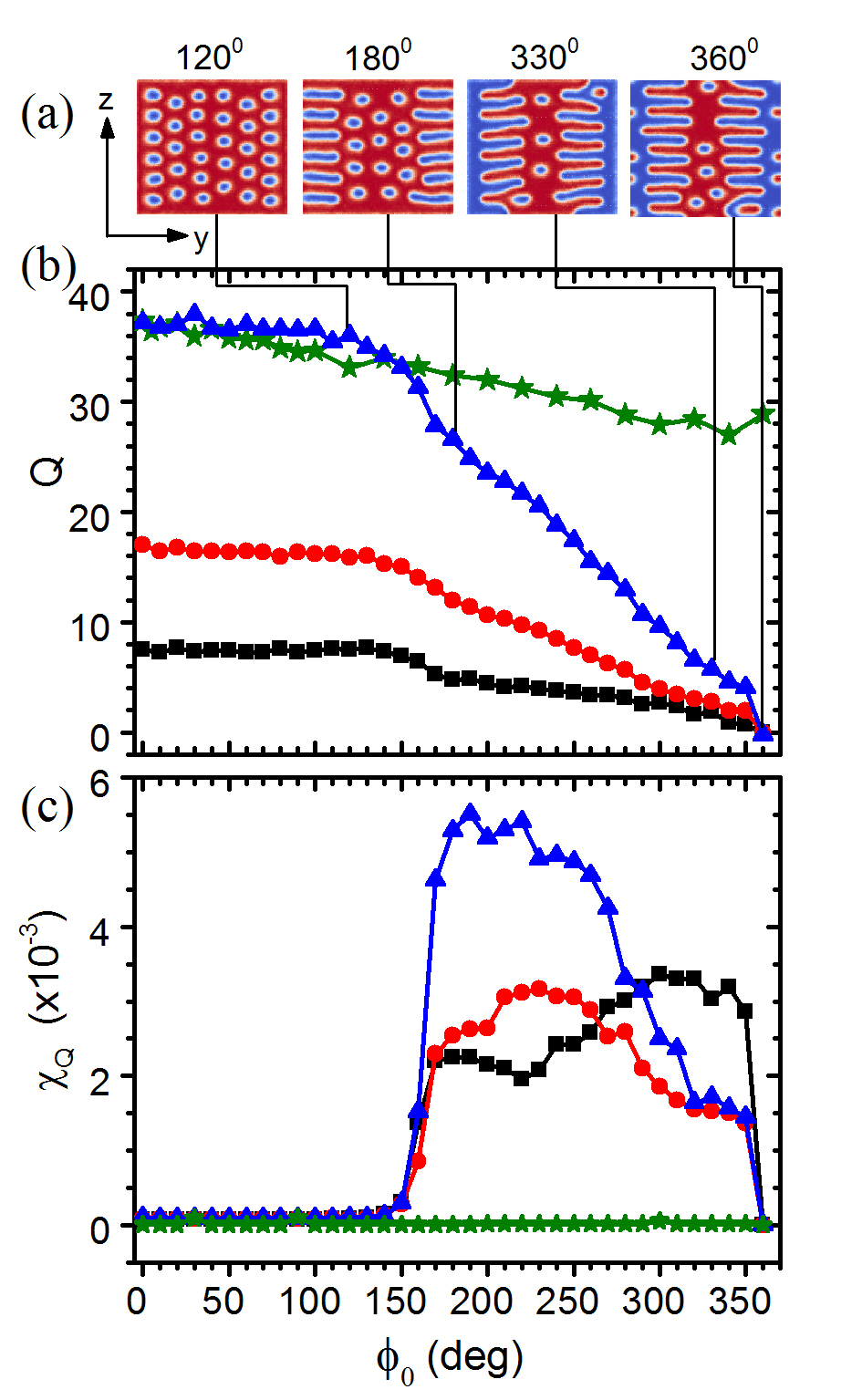}
	\caption{ 
		(Color online) 
		(a) Snapshots of the low temperature magnetic state showing the evolution of the skyrmion phase with curvature angle for a $100a \times 100a$ sample under a uniform magnetic field along the $x$-axis.
		The curved samples are unfoled on the $yz-$plane for visual clarity (see Fig.\ref{fig:sketch}(b)). 
		The color code indicates the projection of local moments on the \textit{radial} direction (red=-1, blue=+1). 
		A nanotube ($\phi_0=360^0$) shows formation of skyrmions on opposite sides with respect to the field direction. 
		These skyrmions have opposite helicities and topological charges.
		(b) Evolution of the topological charge with curvature  angle for different sample sizes:  
		$50a\times50a$ (squares),
		$100a\times50a$ (circles),
		$100a\times100a$ (triangles) and
		$100a\times100a$ under a radial field (stars). 
		(c) Evolution of the topological charge susceptibility with curvature showing a sudden increase at $\phi_0\approx150^0$, associated with the onset of skyrmion deformation close to the lateral sides of the samples.
		Parameters: $d/J=0.5, k/J=0.1, h/J=0.1$ and $k_BT/J=10^{-3}.$
	}
	\label{fig:Q_vs_a}
\end{figure}

The dependence of the topological charge on the curvature angle is shown in  Fig.\ref{fig:Q_vs_a} for nanostrips with different sizes.
We notice that $Q$ remains almost constant up to a characteristic    angle $\phi_0^c\approx 150^0$ and then it drops monotonously to nearly zero for nanotubes, indicating that only a small fraction of the initial number of skyrmions are stable. 

Examination of the low-temperature magnetization distribution in the \textit{radial} direction in Fig.\ref{fig:Q_vs_a}(a) indicates that when a nanotube forms ($\phi_0 \lesssim 360^0$), an equal number of skyrmions in the front ridge (field-out) and the back ridge (field-in) of the nanotube are stabilized. 
By the term ridge we mean here a narrow zone of the cylindrical surface extending parallel to the cylinder axis (z-axis) and containing a generator of the cylinder.
The skyrmions on opposite ridges of the nanotube  have opposite helicity (chirality) and opposite topological charge.
The latter explains the null total topological charge of nanotubes seen in Fig.\ref{fig:Q_vs_a}(b).
The sign inversion of the topological charge and chirality of skyrmions hosted on the front and the back ridge of a nanotube is consistent with the inversion of the field direction relative to the normal to the surface, which in turn is equivalent to chirality inversion of the DMI vectors.

To further characterize the destabilization of the skyrmion phase around the angle $\phi_0^c\approx 150^0$, we compute the topological susceptibility defined as
\begin{eqnarray}
	\chi_Q=(J/k_BT)(<Q^2>-<Q>^2)
	\label{eq:topol_susc}
\end{eqnarray}
and show the results in Fig.\ref{fig:Q_vs_a}(c).
The destabilization of skyrmions and the transformation of
the pure skyrmion phase to the mixed skyrmion-stripe phase 
is identified by a sudden jump in the susceptibility around the characteristic angle $\phi_0^c\approx 150^0$.
We notice however, that this is not a true second order phase transition, namely the susceptibility $\chi_Q$ does not diverge around the characteristic angle.
The gradual decrease of the susceptibility above the characteristic angle  $\phi_0^c$ reflects the coexistence of skyrmions and stripe-like textures in the magnetization distribution.
\linebreak
A final comment is due, regarding the value of the characteristic angle $\phi_0^c$. 
The curving of a planar nanostrip under a uniform applied field, keads to reduction of the normal (radial) component of the field ($h_n$).
For a certain curved nanostrip, the reduction of $h_n$  is more severe near the side edges of the nanostrip and less in the central part (ridge) of the strip, where the field remains almost normal to the surface. 
Therefore, skyrmions initially formed near the side edges of a curved nanostrip deform first (see Fig.\ref{fig:Q_vs_a}(a)) and this occurs when the radial component of the field on the edge of the nanostrip drops below the critical field value.
The phase diagram  (Section~\ref{sec:Sk_PD}) 
predicts for a planar nanostrip with parameters $d/J=0.5, k/J=0.1$ a critical field
$h^c\approx 0.14~d^2/J = 0.035$, below which skyrmions are unstable.
The radial (normal) component of the field on the edge of a curved nanostrip with curvature angle $\phi_0$ is $h_n=  h \cdot sin(\pi/2-\phi_0/2) $ and thus the characteristic angle $\phi_0^c = 2 cos^{-1}(h^c/h) \approx 140^0$.
This result is in reasonable agreement\cite{comm3} with the value of the characteristic angle shown in Fig.\ref{fig:Q_vs_a}.
Thus, the drop of the normal-to-the-surface component of the applied field below the critical value for a planar surface ($h_n < h_c$) is the condition for destabilization of skyrmions on cylindrical nanostrips in a uniform field.

We consider next the case of a  radial applied field with cylindrical symmetry, as in this case the radial component of the field does not change with increasing curvature of the nanostrip.
In Fig.\ref{fig:Q_vs_a}(b) we show that for a radial field the topological charge is weakly dependent on the curvature angle showing $\approx 25\% $ reduction for a nanotube compared to the planar nanostrip.
We attribute the decrease of $Q$ with curvature under a radial field to the non-adaption of the DMI vectors to the surface curvature. 
In particular, under curving of the nanostrip the directions of the DMI vectors around each site remain unchanged and tangential to the surface. 
On the other hand, the applied field, having radial symmetry, adapts to the curvature.
Thus, the total field acting on each site of the lattice is modified due to the local tilting of the applied field.
Consequently, the skyrmion texture of the planar case is perturbed leading to reduction of the topological charge.
The value of $Q$ is expected to decrease monotonously with increasing  curvature and this is actually seen in Fig.\ref{fig:Q_vs_a}(b)

\subsection{Skyrmion shape and size}
\label{sec:Sk_size}
We consider next the evolution of the skyrmion geometrical characteristics, namely size and shape, upon increase of the  nanostrip curvature.
Shape-size analysis of skyrmions in an equilibrium state that hosts an assembly of skyrmions is a numerically intricate task\cite{ziv19} which becomes even more elaborate when the nanostrip is in a mixed skyrmion-stripe phase as it occurs in curved samples  (Figs.(\ref{fig:Skphase}) and (\ref{fig:Q_vs_a})).
To keep the analysis simple, we confine ourselves to nanostrips containing a single skyrmion.
However, to improve the spatial analysis of the magnetization distribution we increase the discretization level of our simulations by reducing the cell size ($a=0.8nm$), but keeping the same material parameters $A_{ex},D,K_u,M_s$ and applied field strength $B$.
This leads to new rationalized parameters
$d/J=0.2$ ($p/a\approx 32$),
$k/J=0.016$ and
$h/J=0.016$.

\begin{figure} [htb!]
	\centering
	\includegraphics[width=0.95\linewidth]{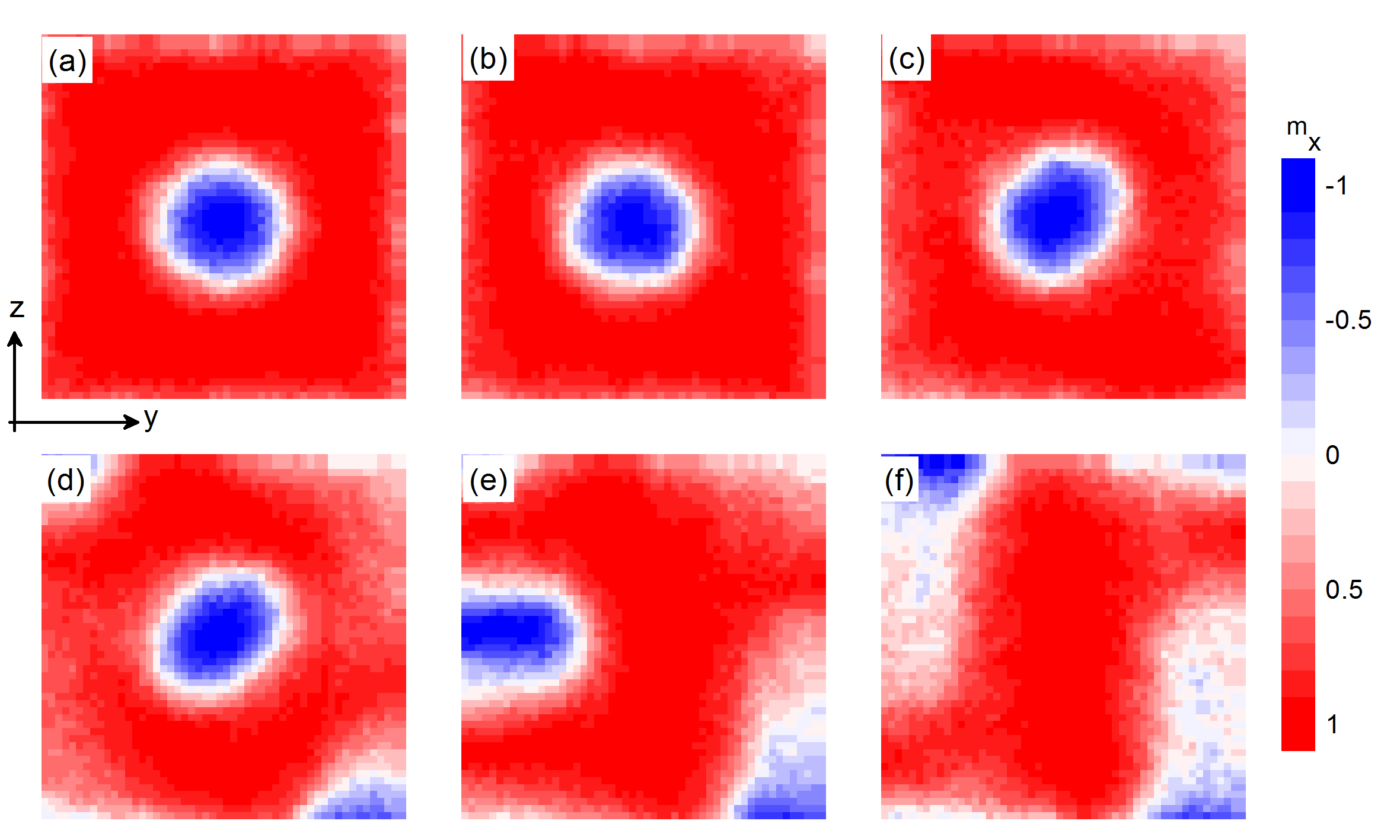}
	\caption{ 
		(Color online) Snapshots of the low temperature magnetization configurations of curved nanostrips hosting a single skyrmion.  Curvature angles and topological charge are
		(a) $\phi_0=  0^0$, $Q=0.84$,
		(b) $\phi_0= 50^0$, $Q=0.83$,
		(c) $\phi_0=100^0$, $Q=0.84$,
		(d) $\phi_0=150^0$, $Q=0.99$,
		(e) $\phi_0=160^0$, $Q=0.52$, and
		(f) $\phi_0=200^0$, $Q=0.10$.
		The curved nanostrips (b)-(f) are unwrapped on the $yz$-plane for visual clarity. 
		A uniform field along the $x$-axis (see Fig.\ref{fig:sketch}(b)) is applied in all cases.
		The color code indicates the values of magnetization  along the field axis.	
		A transformation from purely skyrmion phase (a), to a mixed skyrmion-stripe phase (e,f), due to increasing curvature, is seen.
		System size: $50a\times50a$ $(a=0.8nm)$. Parameters: $d/J=0.2, k/J=0.016, h/J=0.016$ and $k_BT/J=10^{-3}.$
	}
	\label{fig:Single_Sk}
\end{figure}

In Fig.\ref{fig:Single_Sk}, we show the evolution of the single-skyrmion state when it is hosted on nanostrips with gradually increasing curvature.
For small angles ($\phi\lesssim 100^0$), the skyrmion retains its basic geometrical features, such as its size and axially symmetric shape. 
The robustness of the skyrmion at small curvature angles is consistent with the constant value of the topological charge at small curvature angles, seen in  Fig.\ref{fig:Q_vs_a}. 
With increasing curvature, the skyrmion attains a weakly elliptical shape ($\phi \gtrsim 100^0$) and eventually at larger angles ($\phi \approx 160^0$) it annihilates.

\begin{figure} [htb!]
	\centering
	\includegraphics[width=0.95\linewidth]{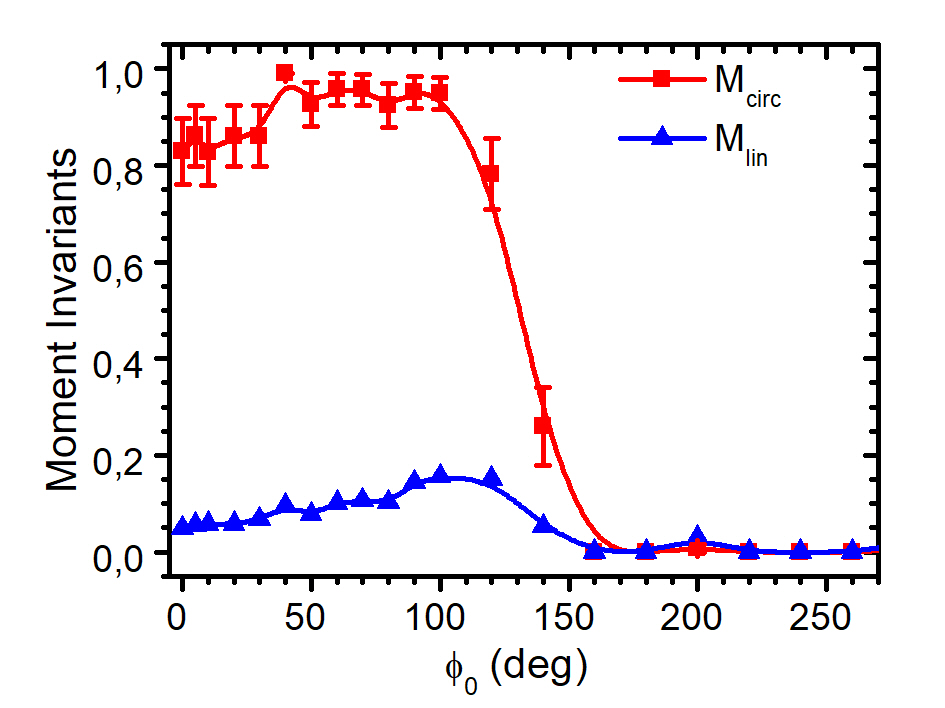}
	\caption{ 
		(Color online) Dependence of skyrmion circularity ($M_{circ}$) and linearity ($M_{lin}$) on curvature angle of a cylindrical nanostrip with size $50a\times50a$ in a uniform applied field. Error bars are obtained from an average over $30$ independent configurations.
		Parameters: $d/J=0.2, k/J=0.016, h/J=0.016$ and $k_BT/J=10^{-3}.$
	}
	\label{fig:Sk_MM}
\end{figure}
To quantify our observations made in Fig.\ref{fig:Single_Sk}, we proceed with a numerical shape analysis of the  isolated skyrmions. 
We define the skyrmion region $S$,
as the compact region of the nanostrip in which the 
local magnetization along the field remains below the saturation value ($m_{i,x}<0.98$ for $B_x>0$) and has topological charge $Q>0.5$.
We compute two shape measures of $S$, namely the invariant moments 
that measure the degree of circularity\cite{zun14} ($M_{circ}$) and linearity \cite{sto08} ($M_{lin} $).
These are defined as
\begin{eqnarray}
	M_{circ}=\frac{\mu_{00}}{\mu_{20}+\mu_{02}} 
	\label{eq:hue1}
\end{eqnarray}
and
\begin{eqnarray}
	M_{lin}=\frac{ \sqrt{(\mu_{20}-\mu_{02})^2+4\mu_{11}^2} }{\mu_{20}+\mu_{02}},
	\label{eq:hue2}
\end{eqnarray}
where the second order geometric moments are
\begin{eqnarray}
	\mu_{pq}=\frac{1}{N_S}\sum_{i\in S}(y_i-y_c)^p(z_i-z_c)^q
	\label{eq:moms}
\end{eqnarray}
with $p,q$ positive integers satisfying $p+q\le 2$, $N_S$ the number of cells in $S$ and $(y_c,z_c)$ the centroid coordinates
$y_c=\sum_i y_i/N_S$ and 
$z_c=\sum_i z_i /N_S$.
In the limiting case of a circular disk the invariant moments are $M_{circ}=1$ , $M_{lin}=0$ and in case of a linear chain  $M_{circ}=0$ , $M_{lin}=1$.

In Fig.\ref{fig:Sk_MM} we show the evolution of the skyrmion shape measures with curvature angle. 
The steep drop of $M_{circ}$ in the range $\phi_0 \simeq 100^0-150^0$ signifies the skyrmion deformation and eventual annihilation. 
Below this characteristic angle, the skyrmion retains to a good approximation the circular shape ($M_{circ}\simeq 1$ and $M_{lin}\simeq 0$).
Taking a closer look at the evolution of the moments with curvature angle two further comments arise. 
First, at small angles $\phi_0\simeq 50^o$ a weak increase of $M_{circ}$ towards unity indicates a closer proximity to the circular shape at intermediate angles and second, a weak hump in $M_{lin}$ around $\phi_0 \simeq 120^o$ indicates weak elongation of the skyrmion before annihilation. 

	A similar elongation of Bloch skyrmions as they approach the annihilation region on cylindrical nanotubes exposed to an external magnetic field has been recently predicted by micromagnetic simulations.\cite{huo19,wan19} 
	The authors \cite{wan19} attributed the deviations from the circular shape to the competition between intrinsic Bloch-type DMI and curvature-induced DMI terms that have different symmetries. 
	In the present study, the elongation of N\'{e}el skyrmions is observed despite the lack of DMI terms with conflicting symmetries and is understood as an intermediate stage in the transformation of skyrmions to stripes.

\begin{figure} [htb!]
	\centering
    \includegraphics[width=0.95\linewidth]{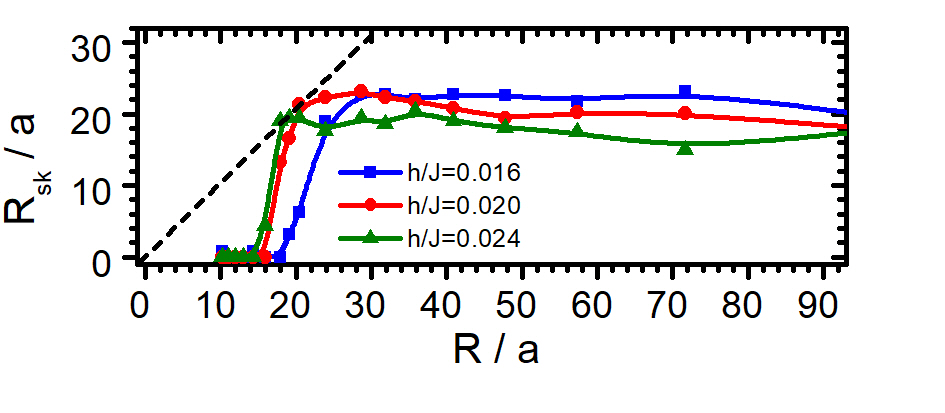}
	\caption{
		(Color online) Dependence of the (effective) skyrmion radius on  curvature radius for nanostrips with size $50a\times50a$ in a uniform applied field. 
		The dashed line is the $R_{sk}=R$ plot that serves as guide to the eye.
		Skyrmion annihilation is observed when $R_{sk} \simeq R$.
		Parameters: $d/J=0.2, k/J=0.016$ and $k_BT/J=10^{-3}.$
	}
	\label{fig:Rsk}
\end{figure}

Before proceeding with the calculation of the skyrmion size we need to clarify some points. 
For deformed (non-circular) skyrmions, as for example, those forming close to the sides of a large nanotubes (Fig.\ref{fig:Q_vs_a}(a)) or on a surface with small curvature radius (Fig.\ref{fig:Single_Sk}), the skyrmion radius cannot be defined in a unique and strict manner.  
A common approach is to fit the elongated skyrmion shape to an ellipse and determine the values of major and minor radii.\cite{huo19}  
For computational efficiency, we have chosen instead to define an effective radius through the relation 
\begin{equation}
R_{eff}=\sqrt{2} R_g,
\label{eq:Reff}
\end{equation}
where $R_g=\sqrt{\mu_{20}+\mu_{02}}$ is the gyration radius that can be computed in a straightforward manner from the Cartesian coordinates of the sites belonging to the skyrmion region $S$.
Then, for a circular skyrmion, the effective radius coincides with the exact radius $(R_{sk}=R_{eff})$ and  
for a non-circular skyrmion Eq.(\ref{eq:Reff})  provides an $rms$ value of the distance distribution from the skyrmion center. 
On the other hand, the shape measures (Fig.\ref{fig:Sk_MM}) indicate that skyrmions remain to a good approximation circular and only close to annihilation they are weakly deformed ($M_{lin}\approx 0.15$ in Fig.\ref{fig:Sk_MM}). 
It is therefore reasonable to approximate the skyrmion radius by the effective radius ($R_{sk}\approx R_{eff}$) for the rest of our study.

In Fig.\ref{fig:Rsk} we show the dependence of skyrmion radius on curvature radius for the same nanostrips as in Fig.\ref{fig:Sk_MM}.
Starting from the planar limit ($R\gg a$), we notice that $R_{sk}$ remains constant as $R$ decreases up to the point that the two radii become approximately equal.
Then a sudden drop of $R_{sk}$ indicates the skyrmion instability and annihilation. 
This behavior is also observed for higher field values ($h/J=0.020, 0.024$), where the skyrmion radius is slightly reduced.  
Seen from the point of view of competing length scales, the curvature radius is a geometrical length while the skyrmion radius a physical length.
Skyrmions are stable when $R/R_{sk} \gg 1$ and instability occurs when  $R/R_{sk} \sim 1$.
This geometrical argument summarizes the instability condition of skyrmions on cylindrical nanostrips as a competition between length scales.
\begin{figure} [htb!]
	\centering
	\includegraphics[width=0.95\linewidth]{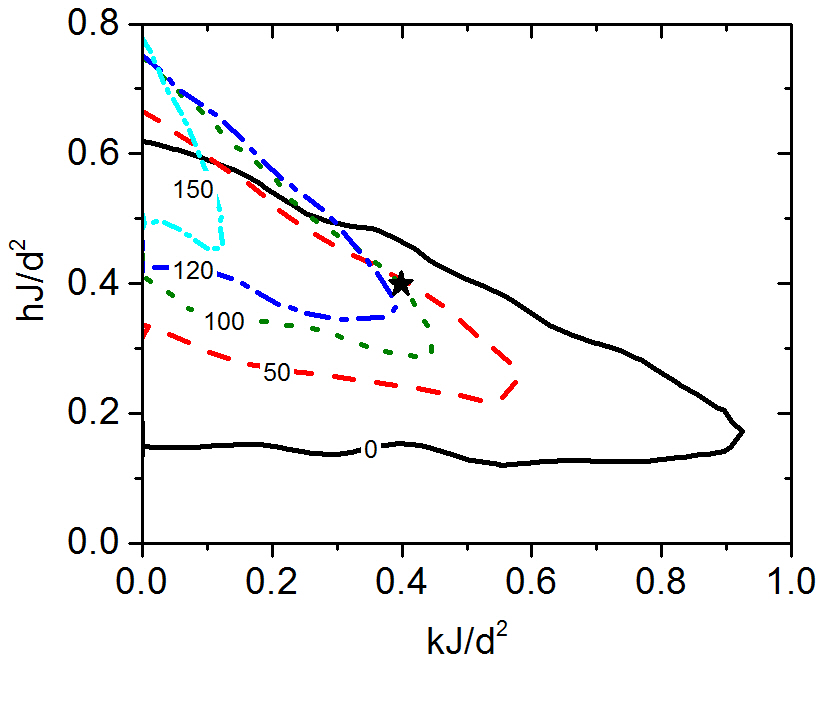}
	\caption{
		(Color online) Scaled field versus scaled anisotropy phase diagram showing the evolution of the skyrmion phase with sample curvature. Lines indicate the boundary of the skyrmion domain, defined for $Q > 0.5$. 
		Curvature angles are indicated on the boundary lines for 
		$\phi_0=0^0$ (solid),
		$\phi_0=50^0$ (dash),
		$\phi_0=100^0$ (dot),
		$\phi_0=120^0$ (dash - dot) and
		$\phi_0=150^0$ (short dash - dot).
		The star indicates the material parameters ($\tilde k =0.4, ~ \tilde h =0.4$) used in the present work. 
		Parameters: sample size $17a \times 17a,~ d/J=0.656$ and $k_BT/J=10^{-3}$. 
		}
	\label{fig:Sk_PD}
\end{figure}
\subsection{Phase Diagram}
\label{sec:Sk_PD}
We conclude this section with the anisotropy-field phase diagram\cite{kee15} under increasing sample curvature. 
The equilibrium magnetization distribution of a nanostip 
is the outcome of a balance between competing energy terms leading to parallelization of the moments (anisotropy, applied field) and orthogonal arrangement of the moments (DMI), both expressed in terms of the exchange energy. 
Thus,  the scaled anisotropy $\tilde{k}=kJ/d^2$ and scaled applied field $\tilde{h}=hJ/d^2$ are the only two dimensionless parameters required to quantify the relative strength of these two competing factors.\cite{comm2}
Notice that the scaled parameters ($\tilde{k},\tilde{h}$) are independent of the discretization level and they depend solely on the material parameters ($A,D,K_u, M_s$). 
We choose the ratio $p/L \simeq 10.8$ equal to the value used in our simulations for systems supporting a single skyrmion (Section~\ref{sec:Sk_size}) in order to have the same finite size effects.
We simulate a field-cooling process for different values of the scaled parameters.
When the topological charge of the final state is $Q > 0.5$ we consider it a skyrmion hosting state.
The resulting lines shown in Fig.\ref{fig:Sk_PD} represent the boundary of the skyrmion domain in the phase diagram.
For a planar nanostrip we reproduce the triangular skyrmion domain discussed previously by Keesman \textit{et al}.\cite{kee15,comm4}
The skyrmion domain shows a gradual shrinkage with curvature and a shift of the skyrmion boundary towards lower anisotropy and higher field values.
The increase of the lower critical field values means that stronger  fields are required to stabilize skyrmions on curved surfaces, because only the radial component of the field assists the stability of skyrmions.
A particular choice of material parameters  is represented by a fixed point in the anisotropy-field diagram. 
The mark (star) in Fig.\ref{fig:Sk_PD} corresponds to the material parameters used in our simulations (see Section~\ref{sec:model}). 
The displacement of the skyrmion boundary relative to the fixed mark 
implies an instability of the skyrmion phase for curvature angles  larger than $\phi_0 \simeq 120^0$, as the material mark lies outside the skyrmion boundary.
This result is consistent with our previously discussed results regarding the suppression of skyrmion circularity (Fig.\ref{fig:Sk_MM}) and skyrmion radius (Fig.\ref{fig:Rsk}) around the same angle. 
A final remark would be that, despite the fact that we do not proceed with a finite size scaling of our results, we anticipate that our findings for the role of curvature will be qualitatively valid for other values of the $p/L$ ratio.
\subsection{Zero-field skyrmions}
\label{sec:Sk_ZF}
\begin{figure} [htb!]
	\centering
	\includegraphics[width=0.95\linewidth]{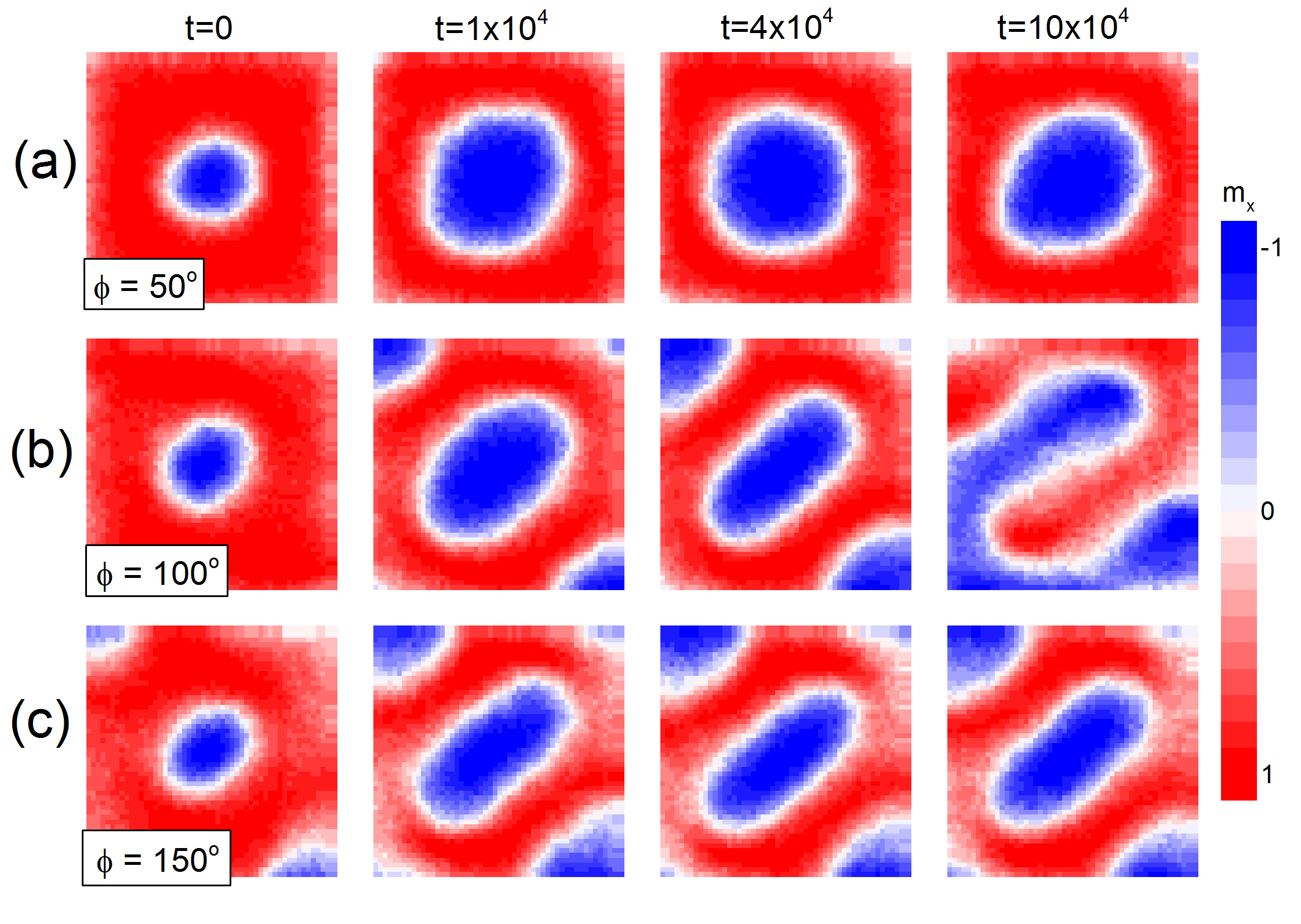}
	\includegraphics[width=0.95\linewidth]{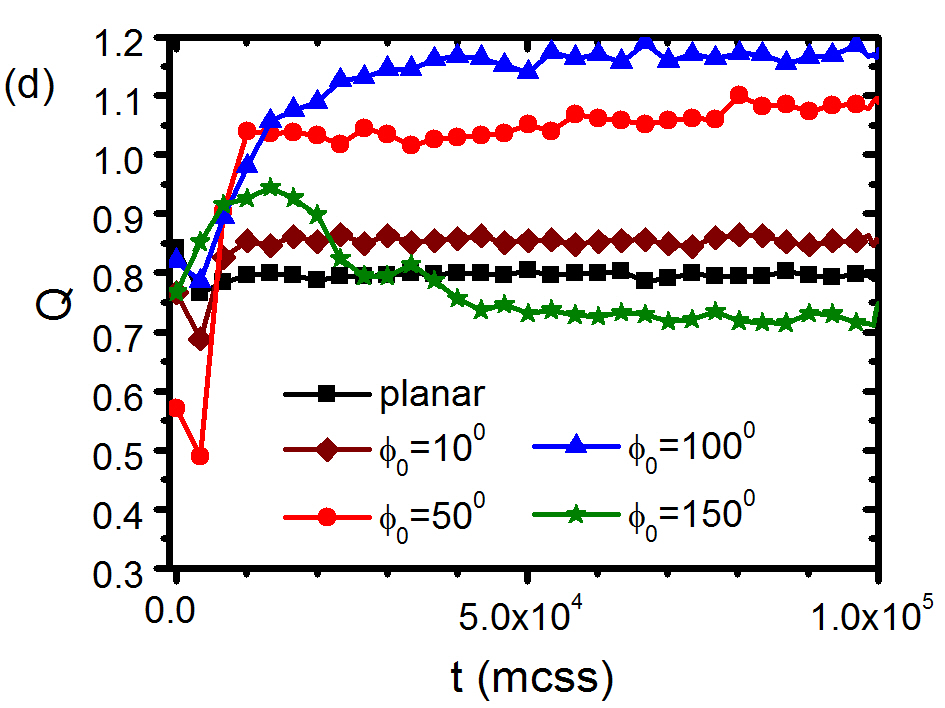}
	\caption{
		(Color online) (a)-(c) Time evolution of zero-field skyrmions in a nanoelement with size $50a\times50a$ and different curvature angles ($\phi_0=50^0,100^0,150^0$). 
        The nanostrips are unwrapped on the $yz$-plane for visual clarity. 
        A uniform field is applied in all cases along the $x$-axis (see Fig.\ref{fig:sketch}(b)).
        The color code indicates the values of magnetization  along the field direction.	
		(d) Time evolution of the topological charge ($Q$) after switching off the applied field ($t=0$)	
		Parameters: $d/J=0.2, k/J=0.016, h/J=0.016$ and $k_BT/J=10^{-3}.$
	}
	\label{fig:Sk_t}
\end{figure}
It has long been established\cite{roh13} that magnetic skyrmions can be stabilized in planar nanoelements of circular shape (dots) in the absence of an applied field, commonly referred to as zero-field skyrmions.
We examine here the possibility of stabilizing zero-field skyrmions in cylindrical nanoelements and study the geometrical limits of stability.
The size of the nanoelement and the skyrmion pitch are chosen, as in the previous section, so that a single skyrmion is stabilized in the nanoelement. 
Free boundaries are assumed in both directions.
To generate a skyrmion, we field-cool the system to low temperature under a uniform field normal to the nanostrip.
Then we switch off the magnetic field and record the time evolution of the magnetization configuration and the topological charge.
To reach the long-term behavior of the system the observation time after switching off the field is 20 times longer (MCSS$ = 10^5$) than the relaxation time used during the field-cooling process (MCSS$ = 0.5 \times 10^3$).
Results for the zero-field relaxation of skyrmions and their topological charge are shown in Fig.\ref{fig:Sk_t}. 
Distinct behaviors are recored for systems with different degree of curvature. 
In case of a planar nanoelement the topological charge remains almost constant in time indicating the stability of skyrmion at zero field.
In systems with small curvature angle  ($\phi_0 \lesssim 100^0$), the skyrmion is still stable, however, its size increases slightly in the absence of a magnetic field, because the Zeeman energy acted in favor of ferromagnetic order and shrinkage of the skyrmion region. 
As seen in Fig.\ref{fig:Sk_t}a, the curvature of the nanoelement enhances the expansion of the skyrmion after switching off the field. 
The weak increase of the topological charge from $Q\simeq 0.8$ to  $Q\simeq 1.2$ that accompanies the increase in size of the zero-field skyrmion ($\phi_0 \lesssim 100^0$) is understood as an outcome of thermal fluctuations and misalignment of the moments along the free boundaries.\cite{roh13}
For larger curvature angles ($\phi_0 \gtrsim 100^0$) the zero-field skyrmion becomes unstable and gradually transforms to a stripe-like structure. 
This behavior is characterized by decreasing values of the topological charge with time. 
In case of planar nanoelemets the stabilization of zero-field skyrmions is attributed to the presence of free boundaries that repel the skyrmion.\cite{roh13}
It becomes clear form Fig.\ref{fig:Sk_t} that the same argument holds in case of a curved nanoelements provided the curvature angle remains below a characteristic angle ($\phi_0 \sim 100^0$) that corresponds to a curvature radius $R/a \sim 28$) close to the zero-field skyrmion radius, Fig.\ref{fig:Sk_t}(a).
\section{Conclusions and Discussion}
We have studied the influence of curvature on the stabilization of nanometer size N\'{e}el skyrmions in thin cylindrical nanostructures with competing Heisenberg  and  Dzyaloshinskii-Moriya exchange interactions.
We showed that application of a uniform magnetic field normal to the cylinder axis could stabilize skyrmions under two conditions.
First, the radial component of the applied field must exceed the critical field for skyrmion formation on the corresponding planar nanostrip, and
second, the curvature radius of the nanostrip must at least exceed the skyrmion radius ($R\gtrsim R_{sk}$). 
These conditions control the shrinkage of the skyrmion-phase pocket in the anisotropy-field phase diagram, under curving of the hosting nanostrip.
Similarly, zero-field skyrmions can also be stabilized on cylindrical nanoelements, provided the above geometrical conditions are satisfied.
In cylindrical nanostrips and nanotubes with large curvature radius ($R > R_{sk}$) and subject to a uniform applied field normal to the cylinder axis  both skyrmion and strip-like phases coexist, which are however spatially separated.
Skyrmions form on the ridge of the curved surface, namely a zone parallel to the cylinder axis where the external field is normal or almost normal to the surface and stripes form on the lateral side of the surface, where the magnetic field is parallel or almost parallel to the surface. 

	A  remark regarding our theoretical model is due.
	Extending the lattice spin model of Eq.(\ref{eq:energy}) by curvature-induced DMI and anisotropy interactions\cite{kra16} is not expected to change qualitatively our results. 
	These terms in conjunction with long-range magnetostatic interactions would improve the numerical accuracy of the critical parameters for skyrmion stability on nanotubes, as for example, the curvature angle (Fig.\ref{fig:Q_vs_a}) and the skyrmion phase boundary (Fig.\ref{fig:Sk_PD}).

From the point of view of physical systems  and their technological applications, composite magnetic nanowires with heavy metal core and thin transition metal shell could be candidate physical systems to support interface skyrmions in the shell layer.
The spatial separation of skyrmions from stripes in the thin shell of these hybrid nanostructures is anticipated to bring new perspectives in current-driven dynamics of skyrmions in cylindrical  nanostructures, since the applied uniform magnetic field on the curved cylindrical shell establishes the required confining energy barrier that holds skyrmions along the ridge of the nanotube and prohibits boundary annihilation.
A recent numerical study of current-driven Bloch skyrmions on cylindrical nanotubes in a uniform applied field normal to the cylinder axis demonstrated this effect.\cite{huo19}
This is anticipated by the fact that Bloch skyrmions on nanotubes of B20 materials (MnSi, etc) are expected to exhibit similar static properties to the N\'{e}el skyrmions studied here.
The case of an applied field with radial symmetry is particularly interesting. 
Our simulations indicate that a cylindrical nanotube in a radial field supports a \textit{pure} N\'{e}el skyrmion phase for any nanotube radius ($R>R_{sk}$), however with weak deformation as the curvature or the field strength increases. 
Despite the fact that the realization of magnetic fields with cylindrical symmetry and curvature radius in the nanoscale is practically unfeasible at present, potential systems, such as magnetic monopoles and nanoscale ferromagnetic needles, have been discussed in the literature.\cite{car15}   
Furthermore, even before achieving magnetic fields with full radial symmetry at the nanoscale, a narrow radial distribution of the applied field around a central direction is expected to widen the width of the region along a nanotube that can host skyrmions.

We believe that our results could stimulate experimental studies of magnetic skyrmions in hybrid nanowires composed of a heavy metal core and a ferromagnetic shell  and  in nanotubes of non-centrosymmetric materials. 
\section*{Acknowledgments}
\noindent 
DK acknowledges helpful discussions with Stavros Komineas.
DK and LT acknowledge financial support by the Special Account for Research of ASPETE through project \textit{NanoSky} (No 80146).
AP is co-financed by Greece and the European Union (European Social Fund- ESF) through the Operational Programme «Human Resources Development, Education and Lifelong Learning» in the context of the project “Strengthening Human Resources Research Potential via Doctorate Research” (MIS-5000432), implemented by the State Scholarships Foundation (IKY).


\begin{thebibliography}{46}
\bibitem[{com({\natexlab{a}})}]{comm3}
	\bibinfo{note}{A discrepancy between the critical curvature angle shown in
		Fig.\ref{fig:Q_vs_a} ($L/a=50, 100$) and the critical angle predicted from
		the the phase diagram in Fig.\ref{fig:Sk_PD} ($L/a=17$) is attributed to
		finite size effects arising from the different size of the simulated
		nanostrips.}
\bibitem[{com({\natexlab{b}})}]{comm2}
	\bibinfo{note}{If non-local magnetostatic terms are non-negligible, as for	example in thick nanostrips, their relative strength would introduce a third factor determining the equilibrium magnetization distributions.}
\end{thebibliography}

\end{document}